\documentclass[twocolumn,showpacs,floatfix]{revtex4} [12/03/2010]

\usepackage{amsmath,amssymb,amstext}
\usepackage{textcomp}
\usepackage{gensymb}
\usepackage{marvosym}
\usepackage{graphicx}
\usepackage{color}

\newcommand{\ang}{$\buildrel _\circ \over {\mathrm{A}}$}  
\newcommand{\sm}{SmFeAsO}
\newcommand{\smxfe}{SmFeAsO$_{1-x}$}
\newcommand{\smfe}{SmFeAsO$_{0.85}$}
\newcommand{\rfe}{$R$FeAsO$_{1-x}$F$_x$}
\newcommand{\dt}{$\Delta^*(T)$}
\newcommand{\dtc}{$\Delta^*(T_c)$}
\newcommand{\dk}{$\Delta^*(T_c)\,/\,k_B$}
\newcommand{\dktc}{$2\Delta^*(T_c)\,/\,k_BT_c$}
\newcommand{\sig}{$\sigma '(T)$}
\newcommand{\ttc}{$T\,/\,T_c$}

\begin{document}

\title{\textbf{Possibility of local pair existence in optimally doped \smxfe\ in pseudogap regime}}

\author{A.\,L.\,Solovjov$^1$\footnote{{}Electronic address:\: solovjov$@$ilt.kharkov.ua}, V.\,N.\,Svetlov$^1$, V.\,B.\,Stepanov$^1$, S.\,L.\,Sidorov$^2$, V.\,Yu.\,Tarenkov$^2$, A.\,I.\,D'yachenko$^2$, and A.\,B.\,Agafonov$^3$}

\affiliation{$^1$B.\,Verkin Institute for Low Temperature Physics \& Engineering National Academy of Sciences of Ukraine, Lenin Ave.~47, 61103 Kharkov, Ukraine}

\affiliation{$^2$A.\,Galkin Institute for Physics \& Engineering National Academy of Sciences of Ukraine, R.\,Luxemburg~72, 83114 Donetzk, Ukraine}

\affiliation{$^3$Institut f\"{u}r Festk\"{o}rperphysik, Leibniz Universit\"{a}t Hannover, Appelstra\ss e~2, D-30167 Hannover, Germany}

\date{\today}

\begin{abstract}

 We report the analysis of pseudogap $\Delta^*$ derived from resistivity experiments in FeAs-based superconductor \smfe, having a critical temperature $T_c$ = 55~K.
 Rather specific dependence \dt\ with two representative temperatures followed by a minimum at about 120~K was observed.
 Below $T_s \approx$  147~K, corresponding to the structural transition in \sm, \dt\ decreases linearly down to the temperature $T_{AFM}\approx$ 133~K. This last peculiarity can likely be attributed to the antiferromagnetic (\textit{AFM}) ordering of Fe spins.
 It is believed that the found behavior can be explained in terms of Machida, Nokura, and Matsubara (MNM) theory developed for the AFM superconductors.

\end{abstract}



\maketitle

%
\section{Introduction}
%

Despite considerable efforts devoted to the study of superconducting pairing mechanism in the new FeAs-based high-$T_c$\ superconductors (HTS's) its physical nature still remains uncertain.
However, there is a growing evidence that it is of presumably magnetic type and all members of the iron arsenide $R$\,FeAsO$_{1-x}$F$_x$\ family, where $R$\ is a lanthanide, are characterized by the long-range (non-local) magnetic correlations \cite{1}.
It is well known that upon electron or hole doping with F substitution at the O site \cite{2,3,4}, or with oxygen vacancies \cite{5,6} all properties of parent $R$FeAsO compounds drastically change and evident antiferromagnetic (\textit{AFM}) order has to disappear \cite{7,8,9}.
However, the comparison with the present \smxfe\ superconductors points towards an important role of low-energy spin fluctuations.
They emerge on doping away from the antiferromagnetic state which is of spin-density wave (SDW) type \cite{10,11}.
Thus, below $T \sim$ 150~K the \textit{AFM} fluctuations, being likely of spin wave type, are believed to noticeably affect the properties of \rfe systems \cite{1,10,11}.
As shown by many studies \cite{10,11,12,13} the static magnetism persists well into the superconducting regime of ferropnictides.
Besides, it was recently shown theoretically that antiferromagnetism and superconductivity can coexist in these materials only if Cooper pairs form an unconventional, sign-changing state \cite{1,13,14}.

\indent In \smxfe\ strongly disordered but static magnetism and superconductivity both are found to exist in the wide range of doping, and prominent low-energy spin fluctuations are observed up to the highest achievable doping levels where $T_c$\ is maximal \cite{10}.
The analysis of the muon asymmetry \cite{11} demonstrates that the coexistence of magnetism and superconductivity  must be nanoscopic, i.\,e., the two
phases must be finely interspersed over a typical length scale of few nm.
Recently reported results on peculiar magnetic properties of LaFeAsO$_{0.85}$F$_{0.1}$\ at $T_{AFM} \approx$ 135~K \cite{15} are likely due to this two-phase structure.

\indent The relation between the SDW and superconducting order is a central topic in the current research on the FeAs-based high-$T_c$\ superconductors.
However, the clear nature of the complex interplay between magnetism and superconductivity in FeAs-based HTS's is still rather controversial.
As a result, rather complicated phase diagrams for different FeAs-based high-$T_c$\ systems \cite{4,12,13,14} and especially for \smxfe\ \cite{3,16,17,18,19} are reported.
For all these HTS's rather wide temperature region is found in which the superconductivity coexists with SDW regime.
For \smxfe\ in a zero magnetic field this temperature region ranges from approximately $x$ = 0.1 up to $x$ = 0.18  \cite{3,10}.
As a result, rather peculiar normal state behavior of the system upon T diminution is expected in this case \cite{3,12,13,14} when $x$ is, let's say, 0.15, as it is in our sample \cite{20}.

\indent To shed more light on the problem in our previous study \cite{20} the fluctuation conductivity (FLC) and \dt, referred to as the pseudogap (PG), derived from resistivity measurements on \smfe\ polycrystal with $T_c \approx$ 55 K were analyzed.
As expected, the temperature behavior of FLC was found to be rather similar to that observed for YBCO films with nearly optimal oxygen doping \cite{21}, whereas \dt\ has demonstrated several peculiar features \cite{20}.
In this contribution we venture to explain found \dt\ peculiarities in terms of Machida-Nokura-Matsubara (MNM) \cite{22} theory developed for \textit{AFM} superconductors as well as to compare the results with Babaev-Kleinert (BK) theory \cite{23} considering superconductors with different charge carrier density $n_f$.

%
\section{Results and discussion}
%

To begin with the pseudogap analysis at first the FLC in \smfe\ polycrystal with $T_c \approx$ 55~K has been thoroughly analyzed \cite{20}.
The FLC is a part of a common excess conductivity $\sigma '(T) = \sigma(T) - \sigma_N(T)$\ which is usually written as
\begin{equation}
\sigma '(T) = [\rho_N(T) - \rho(T)]\,/\,[\rho_N(T) \cdot \rho(T)].
\label{eq:results:sigma-t}
\end{equation}

\noindent Here $\rho(T) = \rho_{xx}(T)$\ is the measured resistivity, and $\rho_N(T) = 1\,/\,\sigma_N(T) = aT + b$\ determines the resistivity of a sample in the normal state extrapolated towards low temperatures.
At the PG temperature $T^*$ = (175\,$\pm$,1)~K the longitudinal resistivity $\rho_{xx}(T)$\ demonstrates a pronounced downturn from its linear dependence at higher temperatures, thus resulting in the excess conductivity.

\indent The common excess conductivity \sig\ as a function of the reduced temperature which is defined as $\varepsilon = ln(T\,/\,T_c^{mf}) \approx (T - T_c^{mf})\,/\,T_c^{mf}$\ is plotted in Fig.~3 and 4 (see Ref. \cite{20}) in a double logarithmic scale.
Here $T_c^{mf} \approx$ 57~K is the mean-field critical temperature \cite{21}.
It was shown that the conventional fluctuation theories by Aslamasov-Larkin (AL) \cite{24} and Hikami-Larkin (HL) \cite{25} well fit the experimental data in the temperature region relatively close to $T_c$.
The result suggests the interband pairing mechanism as a dominant one in \smfe, as it was theoretically discussed in Ref. \cite{26}.
It should be also noted that in the HL theory only the Maki-Thompson (MT) fluctuation contribution was used \cite{20}.

\indent The MT-AL (2D-3D) crossover is distinctly seen in the \sig\ dependence as T approaches $T_c$\ \cite{20}.
Taking into account the crossover temperature $T_0 \approx$ 58.5~K and the distance between \textit{As} layers in conducting \textit{As-Fe-As} planes
$d \approx$ 3.05~\ang, the coherence length along the c-axis $\xi_c(T) = d\,\varepsilon_{c0}^{1/2} =$ (1.4\,$\pm$\,0.005)~\ang\ was determined \cite{20}.
The coherence length $\xi_c(T)$\ is an important parameter of the PG analysis \cite{21}.

%
\subsection{Pseudogap analysis}
%

To analyze PG we assume that the excess conductivity \sig\ at the temperatures $T_c^{mf} \ll T \ll T^*$\ arises as a result of the paired fermions organization in the form of the local pairs (strongly bound bosons (SBB)) \cite{21,27} which satisfy the Bose-Einstein condensation (BEC) theory \cite{28,29,30,31,32}.
Upon temperature decrease the local pairs transform into fluctuating Cooper pairs as $T$\ approaches $T_c^{mf}$\ \cite{21}.
The conventional fluctuation theories describe experiment only up to the temperature $T_{c0} \approx$ 69~K \cite{20}.
Unfortunately, there is still no fundamental theory to describe the experimental curve in the whole PG region.
Nevertheless, the equation for \sig\ has been proposed in Ref. \cite{21} as:
%
%
%
%
%
\begin{equation}
\sigma '(\varepsilon) = \frac{e^2\,A_4\,\left(1 -
\frac{T}{T^*}\right)\,\left(exp\left(-\frac{\Delta^*}{T}\right)\right)}{(16\,\hbar\,\xi_c(0)\,
\sqrt{2\,\varepsilon_{c0}^*\,\sinh(2\,\varepsilon\,/\,\varepsilon_{c0}^*})},
\label{eq:results:sigma-eps}
\end{equation}
\noindent where $A_4$\ is a numerical factor which has the same meaning as a $C$-factor in the FLC theory. In this case the fact is important that Eq.~\ref{eq:results:sigma-eps} contains PG in an explicit form. Besides, the dynamics of pair-creation and pair-breaking below $T^*$\ has been taken into account in order to correctly describe experiment \cite{21}.
To find coefficient $A_4$\ the curve, calculated with Eq.~\ref{eq:results:sigma-eps}, has to fit the $\sigma '(\varepsilon)$\ data in the region of 3D AL fluctuations near $T_c$\ \cite{20,21}.
All other parameters in Eq.~\ref{eq:results:sigma-eps} directly come from resistivity and FLC analysis.
As it was shown in Ref. \cite{20} the curve constructed using Eq.~\ref{eq:results:sigma-eps} with parameters $\varepsilon_{c0}^*$\, =\, 0.616,  $\xi_c(0)$\, =\, 1.405~\ang, $T_c^{mf}$\, =\, 57~K, $T^*$\, =\, 175~K, $A_4$\, =\, 1,98 and \dk\, =\, 160~K describes the experimental data well in the whole temperature interval of interest.

\indent Solving Eq.~\ref{eq:results:sigma-eps} for $\Delta^*$\ we obtain \cite{21}

\begin{equation}
\Delta^*(T) = T\,ln\frac{e^2\,A_4\,(1 - \frac{T}{T^*})}{\sigma '(T)\,16\,\hbar\,\xi_c(0)\,\sqrt{2\,\varepsilon_{c0}^*\,\sinh(2\,\varepsilon\,/\,\varepsilon_{c0}^*)}}.
\label{eq:results:delta-t}
\end{equation}

\noindent Here \sig\ is the experimentally measured value of the excess conductivity in the whole temperature range from $T^*$\ down to $T_c^{mf}$.
All other parameters are the same as designated above. As all the parameters, including \sig, are determined from the experiment, the values of \dt\ can be calculated according to Eq.~\ref{eq:results:delta-t} and plotted now as shown in Fig.~5 of Ref. \cite{20}.

\indent  Unfortunately \dtc\ and in turn the ratio \dktc\ in the FeAs-based HTS's remain uncertain.
At present it is believed that \smxfe\ has two superconducting gaps, i.\,e.\: $\Delta_1(0) \approx$ 6~meV ($\sim$70~K) and $\Delta_1(0) \approx$ (14 \dots 21)~meV ($\sim$160 \dots 240~K) \cite{33}.
Besides we think that \dtc\, $\sim \Delta(0)$\ \cite{21,34}.
That is why four curves are finally plotted in Fig.~5 (Ref. \cite{20}) with \dk\, =\, 160~K (\dktc\, $\sim$\, 5.82 close to strongly coupled limit), 140~K (\dktc\, $\sim$\, 5.0), 120~K (\dktc\, $\sim$\, 4.36) and 100~K (\dktc\, $\sim$\, 3.63 close to weakly coupled BCS limit) from top to bottom, respectively.
Naturally, different values of the coefficient $A_4$\ correspond to each curve whereas the other parameters mentioned above remain unchangeable.

\indent  It was found \cite{20}, that at $T \leq T^*$\ the values of \dt\ start to increase rapidly, as it was observed for YBCO films with different oxygen concentration \cite{21}.
However, an unexpected sharp decrease of \dt\ at $T_s \approx$ 147~Ê was revealed as clearly illustrate Fig.~1 as well as Fig.~5 in Ref. \cite{20}. Usually $T_s$\ is treated as a temperature at which a structural tetragonal-orthorhombic transition occurs in \sm.
In the undoped FeAs compounds it is also expected to be a transition to SDW ordering regime \cite{7,8,9}.
Below $T_s$\ the pseudogap \dt\ drops linearly down to $T_{AFM} \approx$ 133~K, which is attributed to the \textit{AFM} ordering of the Fe spins in a parent \sm\ compound \cite{7,35}.
Below $T_{AFM}$\ the slop of the \dt\ curves apparently depends on the value of \dtc\ \cite{20}.
%
%
%
%
\begin{figure} [t]
\noindent\centering{
\includegraphics[width=0.9\columnwidth]{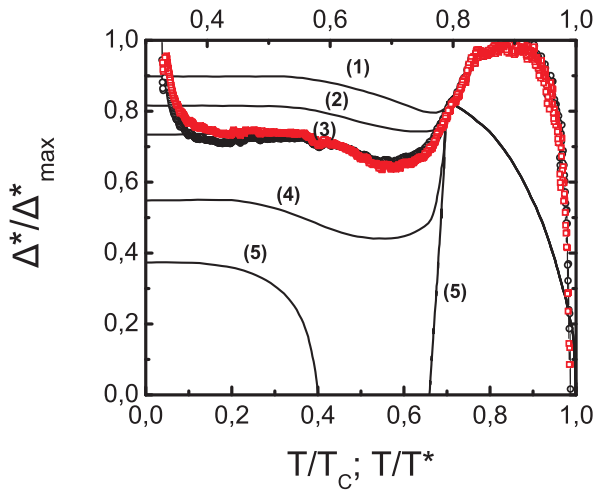}
}
\caption{\dt\,/\,$\Delta_{max}^*$\ in \smfe:\: \textcolor[rgb]{1.00,0.00,0.00}{\Squarepipe}\ \textminus\ \dk\, =\, 130~K; \textcolor[rgb]{0.00,0.00,0.00}{\Circpipe}\ \textminus\  135~K.  Solid curves correspond to MNM theory with the different $\alpha \sim 1\,/\,[\breve{g}\,N(0)]$: (1) \textminus\ $\alpha$\, =\, 0.1, (2) \textminus\ $\alpha$\, =\, 0.2, (3) \textminus\ $\alpha$\, =\, 0.3, (4) \textminus\ $\alpha$\, =\, 0.6 , (5) \textminus\ $\alpha$\, =\, 1.0; $T_N\,/\,T_c$\, =\, 0.7 \cite{22}.}
\label{pic:FigDelta-to-DeltaMax-1}
\end{figure}

\indent Found \dt\ behavior is believed to be explained in terms of the MNM theory (Fig.~\ref{pic:FigDelta-to-DeltaMax-1}) \cite{22} developed for the \textit{AFM} superconductors, in which the \textit{AFM} ordering with a wave vector \textit{Q} may coexist with the superconductivity.
In the MNM theory the effect of the \textit{AFM} molecular field $h_Q(T)$\ ($|\,h_Q\,| \ll \varepsilon_F$) on the Cooper pairing was studied.
It was shown, that below $T_N$\ the BCS coupling parameter $\Delta(T)$\ is reduced by a factor\ [1 - const$\cdot |\,h_Q(T)\,|\,/\,\varepsilon_F$]\ due to the formation of energy gaps of SDW on the Fermi surface along $Q$.
As a result the effective attractive interaction $\breve{g}\,N(0)$\ or, equivalently, the density of states at the Fermi energy $\varepsilon_F$\ is diminished by the periodic molecular field that is

\begin{equation}
\breve{g}\,N(0) = g\,N(0)\,[1 - \alpha m(T)]
\label{eq:results:gn}
\end{equation}

\noindent Here $m(T)$\ is the normalized sublattice magnetization of the antiferromagnetic state and $\alpha$\ is a changeable parameter of the theory. Between $T_c$\ and $T_N$\ ($T_c > T_N$\ is assumed) the order parameter is that of the BCS theory.
Since below $T_N$\ the magnetization $m(T)$\ becomes nonvanishing, $\breve{g}\,N(0)$\ is weakened that results in turn in a sudden drop of $\Delta(T)$\ immediately below $T_N$.
As $m(T)$ saturates at lower temperatures, $\Delta(T)$\ gradually recovers its value with increasing the superconducting condensation energy (Fig.~\ref{pic:FigDelta-to-DeltaMax-1}, solid curves).
This additional magnetization $m(T)$ was also shown to explain the anomaly in the upper critical field $H_{c2}$\ just below $T_N$\ observed in studying of $R$Mo$_6$S$_8$\ ($R$\, =\, Gd, Tb, and Dy) \cite{22}.
However, predicted by the theory decrease of $\Delta(T)$\  at $T \leq T_N$\ was only recently observed in \textit{AFM} superconductor ErNi$_2$B$_2$C with $T_c \approx$ 11~K and $T_N \approx$ 6~K, below which the SDW ordering is believed to occur in the system \cite{36}.
The result evidently supports the prediction of the MNM theory.

\indent Our results are found to be in a qualitatively agreement with the MNM theory as shown in Fig.~\ref{pic:FigDelta-to-DeltaMax-1}, where the data for \dk\,  =\, 130~K (\textcolor[rgb]{1.00,0.00,0.00}{\Squarepipe}) and \dk\,  = 135~K (\textcolor[rgb]{0.00,0.00,0.00}{\Circpipe}) are compared with the MNM theory (solid lines).
The curves are scaled at $T\,/\,T_c$\,  =\, 0.7 and demonstrate rather good agreement with the theory below $T\,/\,T_c$\,  =\, 0.7.
The upper scale is $T\,/\,T^*$.
Both shown \dt\ dependencies suggest the issue that just \dtc\, =\, 133 K would provide the best fit with the theory.
Above $T\,/\,T_c$\,  =\, 0.7 the data evidently deviate from the BCS theory.
It seems to be reasonable seeing SmFeAsO$_0.15$\ as well as any ferropnictide to be not a BCS superconductor.

\indent It is important to emphasize that in our case we observe the particularities of \dt\ in the PG state, i.\,e.\: well above $T_c$\ but just at $T_s$, below which the SDW ordering in parent \sm\ should occur.
It seems to be somehow surprising as no SDW ordering in optimally doped SmFeAsO$_0.15$\ is expected.
On the other hand, the \textit{AFM} fluctuations (low-energy spin fluctuations) should exist in the system as mentioned above.
At the singular temperature $T_s$\ these fluctuations are believed to enhance \textit{AFM} in the system likely in form of SDW.
After that, in accordance with the MNM theory scenario, the SDW has to suppress the order parameter of the local pairs as shown by our results.
Thus, the result suggests the existence of paired fermions in the PG region, which order parameter is apparently suppressed by the \textit{AFM} fluctuations.
These fermions have most likely to exist in the form of the local pairs (SBB) \cite{21}.

\begin{figure} [!t]
\noindent\centering{
\includegraphics[width=0.9\columnwidth]{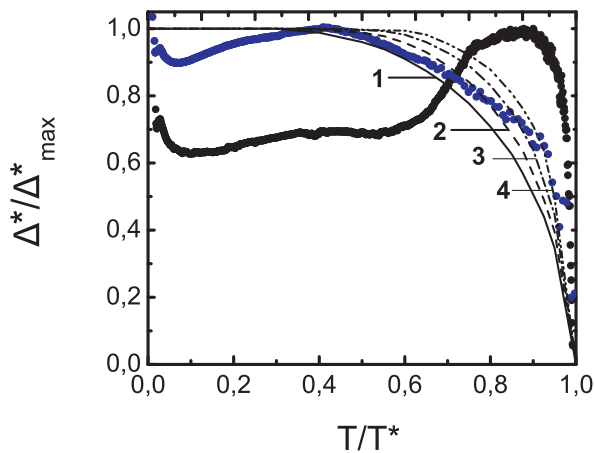}
}
\caption{\dt\,/\,$\Delta_{max}^*$\ in \smfe\ with \dk\, = 160~K (\textcolor[rgb]{0.00,0.00,0.00}{\Circsteel}) and in YBCO film with $T_c$\, =\, 87.4~K (\textcolor[rgb]{0.00,0.00,0.50}{\Circsteel}) \cite{13} as a function of $T\,/\,T^*$\ ($T\,/\,T_c$\ in the case of the theory). Curves 1\,$\dots$\,4\ correspond to BC theory \cite{23} with different $x_0 = \mu\,/\,\Delta(0)$: 1 \textminus\ $x_0$\, =\, 10.0 (BSC limit), 2 \textminus\ $x_0$\, =\, -2.0, 3 \textminus\ $x_0$\, =\, -5.0, 4 \textminus\ $x_0$\, =\, -10.0  (BEC limit).}
\label{pic:FigDelta-to-DeltaMax-2}
\end{figure}

\indent  To justify the issue the relation \dt\,/\,$\Delta_{max}^*$\ as a function of $T\,/\,T^*$\  (\ttc\ in the case of the theory) is plotted in Fig.~\ref{pic:FigDelta-to-DeltaMax-2}.
The dots represent the studied \smfe\ with (\dk\, =\, 160~K.
Blue dots display the data for YBCO film with $T_c$\, = 87.4~K \cite{21}.
The solid and dashed curves display the results of the Babaev-Kleinert (BK) theory \cite{23} developed for the superconducting systems with different charge carrier density $n_f$.
For the different curves the different theoretical parameter $x_0 = \mu\,/\,\Delta(0)$\ is used, where $\mu$\ is the chemical potential.
Curve 1, with $x_0$\, =\, +10, gives the BCS limit.
For curve 2 the value $x_0$\, =\, -2 is taken, for curve 3 parameter $x_0$\, =\, -5.
Finally curve 4 with $x_0$\, =\, -10 represents the BEC limit, which corresponds to the systems with low $n_f$\  in which the
SBB must exist \cite{28,29,30,31,32}.
As well as in YBCO film the \dt\,/\,$\Delta_{max}^*$\  in  \smfe\  evidently corresponds to the BEC limit suggesting the local pairs presence in the FeAs-based superconductor.
Below  \dt\,/\,$\Delta_{max}^* \approx$  0.4 both experimental curves demonstrate the very similar slope suggesting the BEC-BCS transition from local pairs to the fluctuating Cooper pairs found for the YBCO films with different oxygen concentration as temperature approaches $T_c$ \cite{21}.
But, naturally, no drop of \dt\ is observed for the YBCO film (Fig.~\ref{pic:FigDelta-to-DeltaMax-2}, blue dots) as no antiferromagnetism is expected in this case.
This fact accentuates the \textit{AFM} nature of the \dt\ linear reduction below $T_s$\ in \smfe\ found in our experiment.
%
%
%
%
%

%
\section{Conclusion}
%

Analysis of the pseudogap \dt in the FeAs-based superconductor \smfe\ with $T_c$\, =\, 55~K based on the systematic study of the excess conductivity \sig\ \cite{20} has been performed.
Rather specific temperature dependence of the \sig\ was found (Fig.~\ref{pic:FigDelta-to-DeltaMax-1},\ref{pic:FigDelta-to-DeltaMax-2}).
The more striking result is the pronounced decrease of \dt\ below $T_s \approx$ 147~Ê.
As a rule $T_s$\ is treated as a temperature, at which a structural tetragonal-orthorhombic transition occurs in \sm\ \cite{7,8,9}.
In accordance with recent results \cite{3,16,17,18} it is expected to be a transition to SDW ordering regime in the undoped FeAs compounds too.
Below $T_s$\ the pseudogap \dt\ is linear down to $T_{AFM} \sim$ 133~K, which is attributed to the antiferromagnetic ordering of the Fe spins in \sm.
Note that no such peculiarities of $\Delta(T)$\ are observed in the superconducting state of \smxfe\ \cite{37} as no pronounced antiferromagnetism in SC state of the FeAs-based compounds is expected \cite{1,2,3,4,5,6,7,8,9}.

\indent Found \dt\ reduction can be qualitatively explained in the framework of the MNM theory \cite{22}, which predicts the suppression of the superconducting order parameter in AFM superconductors.
But we have to emphasize that we observe the \dt\ reduction in the PG state, i.\,e.\: well above $T_c$.
The finding suggests the presence of paired fermions in \smfe\ in the PG region, the order parameter of which \dt\ is apparently suppressed by the enhanced \textit{AFM} fluctuations (spin waves) in accordance with the MNM theory.
At the same time no unusual drop of \dt\ is observed for the YBCO film (Fig.~\ref{pic:FigDelta-to-DeltaMax-2}) as no antiferromagnetism is expected in this case.
This fact is to justify the \textit{AFM} nature of the found \dt\ reduction in \smfe.

\indent As it is clearly seen in Fig.~\ref{pic:FigDelta-to-DeltaMax-2}, the ratio \dt\,/\,$\Delta_{max}^*$\ in \smfe\ at high temperatures evidently corresponds to the BEC limit.
It seems to be reasonable as in FeAs-based compounds $n_f$\ are found to be at least an order of magnitude less than in conventional metals \cite{17}.
Thus, we may conclude that paired fermions should exist in the PG temperature region of the FeAs-based superconductor \smfe.
Most likely they should appear in the form of local pairs (strongly bound bosons), as it was found for the YBCO films with different oxygen concentration \cite{21}.
Thus, the local pair presence seems to be the common feature of the PG formation in both cuprates and FeAs-based HTS's.

\indent It has to be emphasized that recently reported phase diagrams \cite{3,16,17,18,19} apparently take into account a complexity of magnetic subsystem in SmFeAsO$_{1-x}$F$_x$\ and are in much more better agreement with our experimental results.
But it has also to be noted that we study the \smxfe\ system whereas the phase diagrams are mainly reported for the SmFeAsO$_{1-x}$F$_x$\ compounds.
Is there any substantial difference between the both compounds has yet to be determined.
Evidently more experimental results are required to clarify the question.

\smallskip
%
\begin{center}
\textbf{Acknowledgment}
\end{center}
%

We kindly thank G.\,E.\,Grechnev and Yu.\,G.\,Naidyuk for valuable remarks and discussions.

\end{document}